\documentclass[twocolumn]{aastex631}
\usepackage{acronym,hyperref}

\begin{document}

\title{Improved Upper Limits on Gravitational Wave Emission\\
from NS 1987A in SNR 1987A}

\author{Benjamin J. Owen}
\affiliation{
  Department of Physics and Astronomy,
  Texas Tech University,
  Lubbock, Texas 79409-1051,
  USA
}
\author{Lee Lindblom}
\affiliation{
  Center for Astrophysics and Space Sciences,
  University of California at San Diego,
  La Jolla, California 92093-0424
}
\author{Luciano Soares Pinheiro}
\affiliation{
  Department of Physics and Astronomy,
  Texas Tech University,
  Lubbock, Texas 79409-1051,
  USA
}
\author{Binod Rajbhandari}
\affiliation{
  Department of Mathematics,
  Texas Tech University,
  Lubbock, Texas 79409-1042,
  USA
}
\affiliation{
  Department of Physics and Astronomy,
  Texas Tech University,
  Lubbock, Texas 79409-1051,
  USA
}
\affiliation{
School of Mathematical Sciences and Center for Computational Relativity and
Gravitation, Rochester Institute of Technology, Rochester, NY 14623, USA
}

\begin{abstract}
We report on a new search for continuous gravitational waves from
NS\,1987A, the neutron star born in SN\,1987A, using open data from
Advanced LIGO and Virgo's third observing run (O3).  The search
covered frequencies from 35--1050\,Hz, more than five times the band
of the only previous gravitational wave search to constrain NS\,1987A
[B. J. Owen \textit{et al.,} ApJL \textbf{935,} L7 (2022)].  It used
an improved code and coherently integrated from 5.10\,days to
14.85\,days depending on frequency.  No astrophysical signals were
detected.  By expanding the frequency range and using O3 data,
this search improved on strain upper limits from the previous search
and was sensitive at the highest frequencies to ellipticities of $1.6\times
10^{-5}$ and $r$-mode amplitudes of $4.4\times 10^{-4}$, both an
order of magnitude improvement over the previous search and both
well within the range of theoretical predictions.
\end{abstract}

\acrodef{GW}{gravitational wave}
\acrodef{H1}{Hanford, WA}
\acrodef{L1}{Livingston, LA}
\acrodef{V1}{Cascina, Italy}
\acrodef{O2}{Advanced LIGO and Virgo's second observing run}
\acrodef{O3}{Advanced LIGO and Virgo's third observing run}
\acrodef{O4}{Advanced LIGO and Virgo's fourth observing run}
\acrodef{psd}{power spectral density}
\acrodef{SFT}{Short Fourier Transform}

\section{Introduction}

\citet{Piran1988} first suggested shortly after SN\,1987A that the
neutron star (NS\,1987A) probably born in the supernova could be
emitting detectable continuous \acp{GW}.  Yet the first search to
constrain the behavior of the neutron star through \ac{GW} upper
limits was not performed until recently \citep{Owen_2022}.  That
search used open data from \ac{O2}, covered the frequency band
75--275\,Hz, and was sensitive in that band to \ac{GW} signals just
below an analog of the pulsar spin-down limit based on the age of the
neutron star \citep{Wette2008}.  Here we describe a search of open
data \citep{LVKO3opendata} from \ac{O3} \citep{Tse2019, Acernese2019} using a
new and improved code covering a wider frequency band (35--1050\,Hz).

The detection of neutrinos from SN\,1987A \citep{Bionta1987, Hirata1987}
suggests that a neutron star, rather than a black hole, was the most likely
product of this event.
Located 51.4\,kpc away in the Large Magellanic Cloud
\citep{Panagia1999}, NS\,1987A is the youngest neutron star in our galactic
neighborhood.  Electromagnetic searches for a pulsar or non-pulsing
neutron star in the remnant SNR\,1987A are made difficult by its dust-filled
surroundings.  Far infrared observations of SNR\,1987A by
\citet{Cigan2019}, however, have detected a relatively warm, compact
region of dust that could be powered by a very young cooling neutron
star \citep{Page2020, Dohi2023}.
\citet{Greco2021} and \citet{Greco2022} argue that hard X-ray emission suggests
the presence of a pulsar wind nebula.

The first searches for continuous \ac{GW} emission from NS\,1987A used
stochastic background methods to analyze data from Advanced LIGO and
Virgo's observing runs---see \citet{KAGRA:2021mth} and references
therein.  These searches did not cover a reasonable parameter space.
They assumed spin-down rates for NS\,1987A of order $10^{-9}$\,Hz/s,
which is large by the standards of known pulsars but is at least an
order of magnitude smaller than the spin-down that would be caused by
\ac{GW} emission at detectable levels \citep{Owen_2022}.
Stochastic background searches also did not achieve the needed sensitivity to
detect \acp{GW} at the level of the indirect upper limit on \ac{GW} strain
$h_0^\mathrm{age},$ analogous to the spindown limit for pulsars but based on the
age of the object when pulses are not observed.

This age-based indirect limit was defined by \citet{Wette2008}, who described
the basic continuous-wave method for searches for persistent \acp{GW} from
supernova remnants where there is evidence for a neutron star but where pulses
are not observed (such as SNR\,1987A).
Such searches require that wide bands of frequencies and spin-down parameters
(time derivatives of the frequency) be explored.
These continuous \ac{GW} searches use longer signal coherence times than
stochastic background searches, and therefore generally require searching over
spin-down parameters as well as \ac{GW} frequencies.
The limit $h_0^\mathrm{age}$ is a useful figure of merit for search sensitivity.
The method of \citet{Wette2008} has been used to search for many objects,
starting with the central compact object in supernova remnant Cas~A
\citep{LIGOScientific:2010tql} and more recently targeting NS\,1987A
\citep{Owen_2022}.

Until \citet{Owen_2022}, searches for NS\,1987A did not cover the full parameter
space or reach the sensitivity of $h_0^\mathrm{age}.$
Narrow parameter space searches for continuous \acp{GW} from NS\,1987A were
performed by \citet{Sun2016} using the method described by \citet{Chung2011}.
Since the \citet{Wette2008} parameter space required a fourth spin-down
parameter for NS\,1987A (then 19 years old), \citet{Chung2011} narrowed the
search by introducing a detailed spin-down model.
This is less robust than models making fewer assumptions.
Even with a narrow parameter space, \citet{Sun2016} did not achieve upper limits
comparable to $h_0^\mathrm{age}.$
Recent all-sky surveys for continuous \acp{GW} such as \citet{Dergachev:2022lnt}
do beat that limit in the direction of NS\,1987A, but do not cover spin-down
ranges physically consistent with NS\,1987A.

\citet{Wette2008} derived $h_0^\mathrm{age}$ for mass-quadrupole \ac{GW}
emission and \citet{Owen2010} extended it to current-quadrupole \ac{GW} emission
from $r$-mode oscillations [see \citet{Glampedakis2018} for a summary of
emission mechanisms].
The derivations of these limits assume that \acp{GW} dominate the spindown of
the star from birth and that the initial spin frequency was much higher than
present.
The \citet{Wette2008} mass-quadrupole age limit on the \ac{GW} amplitude $h_0$
[a measure called the intrinsic strain \citep{Jaranowski1998}] can be written as
a frequency-independent expression that depends on the age of the neutron star
$a$, its distance $D$, and moment of inertia $I$:
\begin{eqnarray} 
  h_0^\mathrm{age} &=& \sqrt{\frac{5}{8}}\sqrt{\frac{G I}{c^3 a D^2}}.
  \label{e:Wetteh0age}
\end{eqnarray}
The analogous indirect limit for \ac{GW} emission from $r$-modes is given by
\citep{Owen2010}
\begin{eqnarray}
h_0^\mathrm{age} &=& \sqrt{\frac{10}{A^2(n-1)}}\sqrt{\frac{G I}{c^3 a D^2}},
  \label{e:Owenh0age}
\end{eqnarray}
where $n=f\ddot f/\dot f^2$ is the braking index.
Here we have modified the expression of \citet{Owen2010} to leave free the
parameter $A,$ which is the ratio of $r$-mode frequency (in an inertial frame)
to stellar spin frequency.

We convert the above expressions to numerical ranges as follows.
\citet{Owen2010} used $A=4/3,$ appropriate for slow rotation and
Newtonian gravity, while general relativistic slow rotation estimates
are about 1.39--1.64 \citep{Idrisy2015, Ghosh2023}.  The
moment of inertia depends on the neutron star equation of state and mass.
We take the
mass range for neutron stars to be 1.2--2.1\,$M_\odot$
\citep{Martinez2015, Cromartie2020}.  For this mass range and the
equations of state used by \citet{Abbott_2021}, the moment of inertia
range is about $9.1\times 10^{44}$--$4.7\times 10^{45}$\,g\,cm$^2.$ Finally we use 6--7 for the physically
  plausible range of $r$-mode braking indices \citep{Lindblom1998, Ho2000}.
The resulting range of
$h_0^\mathrm{age}$ for NS\,1987A, using an age of 33 years
(applicable for late \ac{O3}), is about
\begin{equation}
2.3\times 10^{-25} \leq h_0^\mathrm{age} \leq 5.3\times 10^{-25}  
  \label{e:Wetteh0age2}
\end{equation}
for mass-quadrupole \ac{GW} emission using Eq.~(\ref{e:Wetteh0age}) and
\begin{equation}
2.3\times 10^{-25} \leq h_0^\mathrm{age} \leq 6.8\times 10^{-25}  
  \label{e:Owenh0age2}
\end{equation}
for $r$-mode \ac{GW} emission using Eq.~(\ref{e:Owenh0age}).

Inserting these parameters into the results of \citet{Wette2008} and
\citet{Wette2012} indicates that a coherent search of \ac{O3} data using only
two spin-down parameters can surpass the sensitivity of $h_0^\mathrm{age}$ for a
computing budget of order a million core-hours.
This paper describes such a search, which detected no astrophysical signals but
placed direct upper limits on the \ac{GW} strain from NS\,1987A.
These limits beat the indirect limit $h_0^\mathrm{age}$ over a physically
consistent parameter space that is considerably larger than the range of
frequencies explored in the \ac{O2} data search by \citet{Owen_2022}.

\section{Search methods}

The search methods used in this paper are similar to those used by
\citet{Owen_2022}.
Highlights and changes are summarized here.
Readers are directed to \citet{Owen_2022} and references therein for details.

\begin{table*}
  \begin{center}
\begin{tabular}{llll}
\multicolumn{4}{c}{Derived parameters} \\
Name & Value (35--125\,Hz) & Value (125--450\,Hz) & Value (450--1050\,Hz)\\
\hline
Span & 14.85\,d & 8.13\,d & 5.10\,d \\
Start & 2020-02-28 12:39:33 & 2020-02-24 02:56:27 & 2020-02-27 12:34:01 \\
H1 SFTs & 555 & 330 & 216 \\ 
L1 SFTs & 603 & 333 & 220 \\
V1 SFTs & 539 & 263 & 169 \\
\hline
\end{tabular}
\end{center}
\caption{
\label{pars}
Data parameters used in this search.
Times are UTC.
}
\end{table*}

The \texttt{Drill} pipeline \citep{Drill} version 1.0.0 was used for this search.
It will be described more fully elsewhere.
Here we summarize the differences between \texttt{Drill} and previous codes.
\texttt{Drill} is a completely new code with functionality similar to that used
in \citet{Owen_2022}.
It consists of Python scripts running C codes from \texttt{LALSuite}
\citep{LALSuite} (v6.25.1 of \texttt{lalapps} and concurrent versions of other
packages) that implement the multi-detector $\mathcal{F}$-statistic
\citep{Jaranowski1998, Cutler2005}.
\texttt{Drill} is more efficient than the code used in \citet{Owen_2022}, vetoes
signal candidates based on nonparametric statistics, and handles upper limits
more consistently.

This search used open data \citep{Vallisneri2015, LVKO3opendata} from \ac{O3} in
the form of 1800\,s \acp{SFT} generated from Frame-format time-domain data
sampled at 4\,kHz.
All \ac{SFT} logic and data selection were handled by the
\texttt{lalapps\_MakeSFTs} program.
\ac{O3} and our search included data from the \ac{H1} and \ac{L1} 4\,km LIGO
interferometers and the \ac{V1} 3\,km Virgo interferometer.
Although \ac{V1} was generally less sensitive, we verified that including its
data improved search sensitivity at fixed computational cost.
The observation time spans for our search bands were also set to achieve a fixed
computational cost.
Following \citet{Jaranowski1998} the start time of each span was chosen to
maximize the data time divided by the joint \ac{psd} of strain noise, which is
approximately equivalent to maximizing the search sensitivity.

For NS\,1987A we used the (J2000) right ascension and declination
\begin{equation}
\alpha = 05^\mathrm{h}\, 35^\mathrm{m}\, 27^\mathrm{s}.998,
\qquad
\delta = -69^\circ\, 16'\, 11''.107
\end{equation}
from \citet{Cigan2019}.
We used an age of 33\,yr (suitable for the late \ac{O3} data used) and the
distance 51.4\,kpc \citep{Panagia1999} to determine the parameter space and
infer source properties.
Observation spans and other search parameters derived from the age are listed in
Table~\ref{pars}.

The signal parameter space was chosen similarly to \citet{Owen_2022}.
The \ac{GW} frequency in the solar system barycenter frame was modeled by
\begin{equation}
\label{foft}
f(t) = f + \dot f (t-t_0) + \frac{1}{2} \ddot f (t-t_0)^2,
\end{equation}
where $t_0$ is the time at the beginning of the span and the parameters $(f,
\dot f, \ddot f)$ are evaluated at epoch $t_0$.
The ranges of $(\dot f, \ddot f)$ for a given value of $f$ were
\begin{eqnarray}
\frac{f} {(n_{\max}-1)a} \le & -\dot f & \le \frac{f} {(n_{\min}-1)a},\\
n_{\min} \frac{\dot f^2} {f} \le & \ddot f & \le n_{\max} \frac{\dot f^2} {f},
\end{eqnarray}
with the braking index $n$ ranging from $n_{\min}=3$ to $n_{\max}=7$ and $a$
being the neutron star's age.
Unlike in previous searches, a minimum braking index of $n_{\min}=3$ was used to
keep the highest values of $-\dot{f}$ below about $5\times10^{-7}$\,Hz/s, where
\citet{Owen_2022} found that the \ac{SFT} length of 1800\,s can become
problematic.
We verified that the other consistency checks described in \citet{Owen_2022}
were satisfied.
This range of braking indices is consistent with the minimum spin-down for a
given $h_0$ \citep{Owen2010},
\begin{eqnarray}
&&-\dot f = 1.6 \times 10^{-8} \mbox{Hz/s} \left( \frac{A} {2} \right)^2
\left( \frac{D} {\mbox{51.4 kpc}} \right)^2
\nonumber\\
&& \times \left( \frac{h_0} {2\times10^{-25}} \right)^2 \left( \frac{f}
{\mbox{100 Hz}} \right) \left( \frac{10^{45}\,\mbox{g cm}^2} {I} \right)
\end{eqnarray}
and is greater than the maximum value covered by the all-sky surveys such as
\citet{Dergachev:2022lnt}.
It does not include some of the more extreme range recently proposed by
\citet{Morales2023}, which is more appropriate for older stars.

The frequency band was split into a low frequency band from 35--125\,Hz, a
medium frequency band from 125--450\,Hz, and a high frequency band from
450--1050\,Hz.
The boundary at 125\,Hz was chosen so that even the fastest spinning young
pulsar \citep{Marshall1998} emits \acp{GW} in the low band.
The boundary at 450\,Hz was chosen so that the middle band would avoid noise
artifacts due to violin modes of the LIGO test mass suspensions
\citep{Covas2018, Davis2021}.
The overall lower bound of 35\,Hz was chosen so that an \textit{a priori}
estimate of search sensitivity \citep{Wette2012} indicated that
Eq.~(\ref{e:Wetteh0age}) would be achieved.
The overall upper bound of 1050\,Hz was chosen for the same reason and because
going much higher in frequency causes difficulties with the analysis such as the
spin-down range mentioned above.

This search ran a total of roughly $1.2\times10^6$ core-hours on the Texas Tech
``Nocona'' computing cluster, split into batch jobs of about 8 core-hours each.
Integration times and other parameters are shown in Table~\ref{pars}.
Each search job covered the full range of
spin-down parameters appropriate for its frequency band.  Widths of
these bands ranged from about 1--37\,mHz depending on frequency.
Parameter spacings in $\dot f$ and $\ddot f$ were of order
$10^{-11}$\,Hz/s and $10^{-18}$\,Hz/s$^2$ respectively.  Search
jobs contained about 4--$9\times10^8$ templates each, for totals of
about 2, 4, and $6\times10^{13}$ templates for the low, medium, and
high frequency bands respectively.

The approach of \texttt{Drill} to vetoing signal candidates does not rely on
time-frequency behavior or known instrumental lines for \textit{a priori}
vetoes, as in \citet{Owen_2022}.
Instead it outputs a $2\mathcal{F}$ histogram for each search job without
recording any specific candidates on the first pass.
This ameliorates major storage and input/output issues in noisy bands.
From each histogram \texttt{Drill} takes the loudest $2\mathcal{F}$ (to within
the binning resolution of 0.1) and computes (approximately due to the binning)
the discrete Cram\'er-von Mises statistic~\citep{Choulakian_1994}.
The continuous version of the Cram\'er-von Mises statistic can be written
\begin{equation}
\omega^2 = \int dC^*(2\mathcal{F}) [C(2\mathcal{F}) - C^*(2\mathcal{F})]^2,
\end{equation}
where $C$ and $C^*$ are the observed and expected cumulative
distributions respectively.
For the $\mathcal{F}$-statistic, the latter is a $\chi^2$ with four degrees of
freedom.
Thus $\omega$ emphasizes the middle of the distribution (where non-Gaussian
noise is concentrated) more than the tail (where a detectable signal will be)
and is a good way of checking for ``bad'' noise bands without discarding loud
signals.

We determined the threshold of $\omega$ empirically.
Even for Gaussian stationary noise, the output of the \texttt{LALSuite}
$\mathcal{F}$-statistic code is not precisely $\chi^2$(4) distributed due to
approximations used to increase computational speed.
We checked the behavior of $\omega$ in real noise and simulated Gaussian noise,
with and without injected signals.
In Gaussian or nearly Gaussian noise, we found that $\omega$ has a mean of
0.0035 and a standard deviation of 0.0012.
Therefore $\omega=0.02$ corresponds to about thirteen standard deviations and
should not veto clean noise bands even with our large number of trials.
Loud enough signals should cause a high $\omega$ due to many templates
triggering at high $2\mathcal{F}$ with slightly wrong parameters.
Through injection studies we found that signals are not spuriously vetoed until
$2\mathcal{F} \sim 10^6,$ which is orders of magnitude above the physical limit
$h_0^\mathrm{age}.$

\section{Search results}

We checked the search results for candidate signals as follows.
First we vetoed the entirety of each search job which produced $\omega \ge
0.02,$ totaling about 20\,Hz or 2\% of the total search band.
We calculated the 95\% confidence threshold in idealized Gaussian noise for each
wide band ($2\mathcal{F}$ about 75.6, 76.5, and 76.4 respectively) and recorded
which jobs exceeded it (4, 6, and 36 jobs respectively).
Then we visually inspected histograms of the surviving jobs by the same criteria
as in \citet{LIGOS6SNRs}, looking for fat tails rather than the thin tails
indicative of injected signals.
All histograms but one were rejected at a glance, and the remaining one (with
$2\mathcal{F}=79.8$ and $\omega=0.001$) hinted at abnormality on closer
inspection.
Also, that job searched around 998.4\,Hz, a frequency known to be contaminated
(in LIGO data) by violin modes of the test mass suspensions~\citep{Covas2018,
Davis2021}.
Nevertheless, we followed it up.
That search job was rerun, keeping detailed information on all templates with
$2\mathcal{F} \ge 40.$
Plotting $2\mathcal{F}$ versus frequency showed multiple peaks indicative of
noise lines.
We then searched the entire frequency band of that job with double the
integration time.
A real signal would produce double the value of $2\mathcal{F},$ or about 160.
The double-length followup search produced several $2\mathcal{F}$ peaks only a
little over 100, and every search job dramatically failed the Cramer-von Mises
test ($\omega \sim 0.05$).
Therefore we concluded that we found no astrophysical signals.

We then set upper limits on $h_0$ in 1\,Hz bands, similar to \citet{Owen_2022}.
We considered a population of signals with fixed $h_0$ and random values of
intrinsic parameters $(f,\dot f,\ddot f)$ as well as extrinsic parameters
described in \citet{Jaranowski1998}.
We estimated what $h_0$ would be detected at a rate of 90\% with $2\mathcal{F}$
larger than the largest non-vetoed search result in that upper limit band.
A semi-analytic estimate of $h_0$ was checked by software-injecting 1000 signals
per 1\,Hz upper limit band.
The \texttt{Drill} pipeline does this more efficiently by cutting down on disk
input/output bottlenecks, corrects a minor inconsistency in the way vetoed bands
were incorporated into the upper limits in previous analyses, and computes upper
limits for all bands that were not too heavily vetoed.
A heavily vetoed band is one where eliminating the search jobs exceeding the
$\omega$ threshold vetoes more than 10\% of the upper limit band, and therefore
a 90\% upper limit is not meaningful.
These amounted to 60 upper limit bands out of the 1015 covered by the search, or
about 6\%.

\begin{figure*}
  \begin{center} 
    \includegraphics[height=0.35\textwidth]{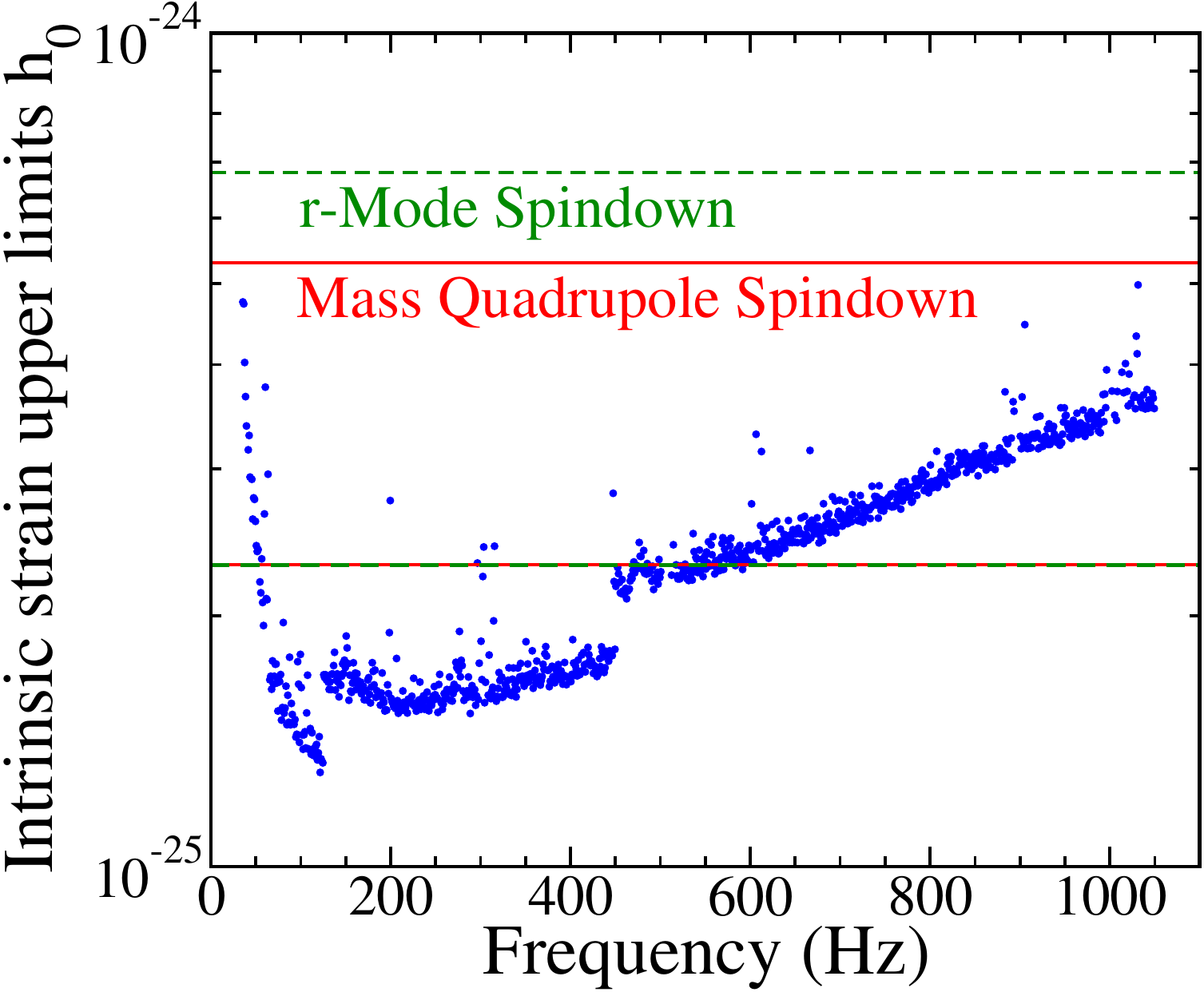}
    \hspace{0.25cm}
    \includegraphics[height=0.35\textwidth]{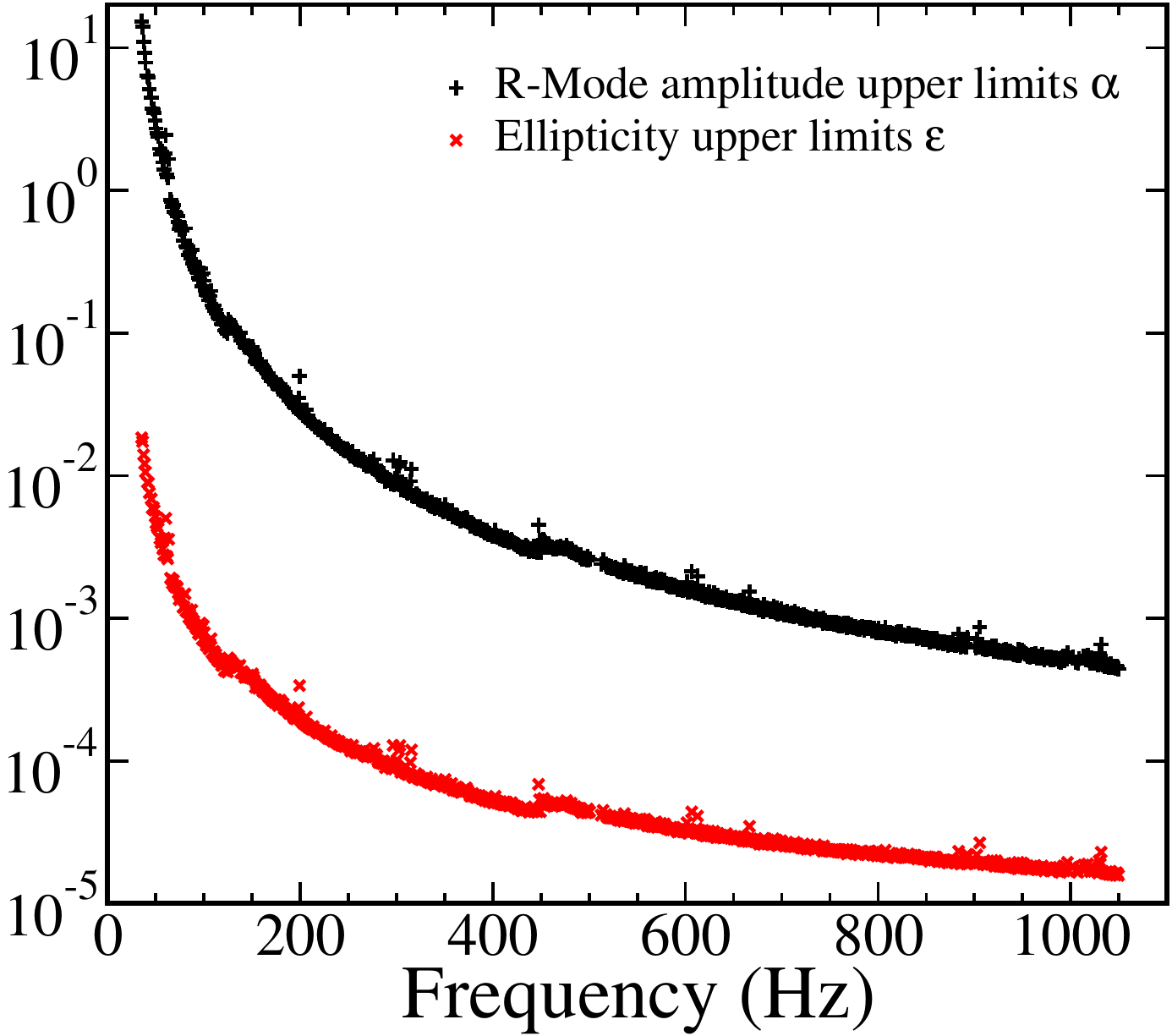}
    \hspace{0.25cm}
  \end{center}
\caption{\label{f:Fig1} The left panel displays observational 90\%
  confidence upper limits on $h_0$ from NS\,1987A
  in 1\,Hz bands as a function of frequency.  The (red) horizontal
  solid lines show the range of $h_0^\mathrm{age}$ from
  mass-quadrupole \ac{GW} emission, Eq.~(\ref{e:Wetteh0age2}), and
  the (green) horizontal dashed lines show the range
  for $r$-mode \ac{GW} emission, Eq.~(\ref{e:Owenh0age2}).
  (Note that the $h_0^\mathrm{age}$ lower limits coincide.)
  The right panel shows corresponding upper limits on the
  dimensionless neutron-star ellipticity $\epsilon$ and $r$-mode
  amplitude $\alpha.$}
\end{figure*}

The left panel of Fig.~\ref{f:Fig1} displays our 90\% confidence upper
limits on $h_0$ as a function of frequency, except in the 60 heavily
vetoed bands.
The discontinuities at 125 and 450\,Hz are caused by differences in
integration times in the three search bands.
The (red) horizontal solid lines in the left panel of Fig.~\ref{f:Fig1} show the
range of $h_0^\mathrm{age}$ from mass-quadrupole \ac{GW} emission,
Eq.~(\ref{e:Wetteh0age2}), and the (green) horizontal dashed lines show the
range for $r$-mode \ac{GW} emission, Eq.~(\ref{e:Owenh0age2}).
Note that the lower red and green lines coincide.
The observed upper limits on $h_0$ are less (better) than the average values of
the indirect limits $h_0^\mathrm{age}$ over the full frequency band.
Our search places limits on \ac{GW} emission from NS\,1987A that are better than
the strictest $h_0^\mathrm{age}$ estimates over the
astrophysically most interesting part of the frequency band, 50--600\,Hz.

The efficiency of our search can be expressed in terms of the sensitivity depth
\citep{Behnke2015}
\begin{equation}
\mathcal{D} = \sqrt{S_h / h_0},
\end{equation}
where $S_h$ is the harmonic mean strain noise power spectral density.
For our search $\mathcal{D}$ is about 29\,Hz$^{-1/2},$ 22\,Hz$^{-1/2},$ and
18\,Hz$^{-1/2}$ in the low, middle, and high frequency bands respectively.
This is comparable to or somewhat worse than \citet{Owen_2022}, as one would
expect from the short integration times \citep{Wette2023}.

Upper limits on $h_0$ imply upper limits on the fiducial neutron star
ellipticity $\epsilon$ \citep{Jaranowski1998}
\begin{equation}
\epsilon \simeq 9.5\times10^{-5} \left( \frac{h_0} {1.2\times10^{-24}} \right)
\left( \frac{D} {\mbox{1 kpc}} \right) \left( \frac{\mbox{100 Hz}} {f}
\right)^2,
\end{equation}
and upper limits on the $r$-mode amplitude $\alpha$ defined by
\citet{Lindblom1998} using \citet{Owen2010},
\begin{equation}
\alpha \simeq 0.028 \left( \frac{h_0} {10^{-24}} \right) \left( \frac{\mbox{100
Hz}} {f} \right)^3 \left( \frac{D} {\mbox{1 kpc}} \right).
\end{equation}
The fiducial values in these expressions are uncertain by roughly a factor of
three due to uncertainties in the neutron star mass and equation of state.
Moreover, general relativity complicates these expressions in ways that have
not yet been calculated.
These upper limit estimates on $\epsilon$ and $\alpha$ for NS\,1987A from this
search are shown in the right panel of Fig~\ref{f:Fig1}.

\section{Conclusions}

We have performed a search for continuous \acp{GW} from NS\,1987A over a much
wider range of frequencies than the only previous physically consistent search
\citep{Owen_2022} using an improved code on improved data.
While we did not detect any astrophysical signal, we set upper limits which
constrained the behavior of NS\,1987A over a much broader parameter space than
ever before.

When translated into fiducial neutron star ellipticity or
$r$-mode amplitude, our upper limits (at the highest frequencies)
are an order of magnitude better than the highest frequency limits
set by \citet{Owen_2022}.  Our best ellipticities are just above
$10^{-5}.$ Such elastic deformations are possible for quark stars and
quark-baryon hybrid stars \citep{Owen2005, Johnson-McDaniel2013}, and are
skirting the maximum predicted for normal neutron stars
\citep{Morales2022}.  For a magnetic deformation, our strain upper
limits imply a limit on the internal magnetic field of a few times $10^{15}$ or
$10^{14}$\,G if the protons in the core are or are not superconducting
respectively \citep{Lander2014, Ciolfi2013}.  Our best upper limits on $r$-mode
amplitude are starting to enter the range of theoretical predictions
\citep{Bondarescu2009}.

These comparisons show that our search had a sensitivity to
\ac{GW} emission from NS\,1987A compatible with a variety of theoretical
predictions, not just the most extreme ones as in \citet{Owen_2022}.
We used simple coherent integrations of relatively short spans of \ac{O3} data.
Searches of better data from the current \ac{O4} to Cosmic Explorer
\citep{Evans2021, Gupta2023} will further improve on this sensitivity,
especially with more sophisticated data analysis techniques \citep{Wette2023}.

\begin{acknowledgments}

This research has made use of data or software obtained from the Gravitational
Wave Open Science Center (gwosc.org), a service of the LIGO Scientific
Collaboration, the Virgo Collaboration, and KAGRA. This material is based upon
work supported by NSF's LIGO Laboratory which is a major facility fully funded
by the National Science Foundation, as well as the Science and Technology
Facilities Council (STFC) of the United Kingdom, the Max-Planck-Society (MPS),
and the State of Niedersachsen/Germany for support of the construction of
Advanced LIGO and construction and operation of the GEO600 detector. Additional
support for Advanced LIGO was provided by the Australian Research Council. Virgo
is funded, through the European Gravitational Observatory (EGO), by the French
Centre National de Recherche Scientifique (CNRS), the Italian Istituto Nazionale
di Fisica Nucleare (INFN) and the Dutch Nikhef, with contributions by
institutions from Belgium, Germany, Greece, Hungary, Ireland, Japan, Monaco,
Poland, Portugal, Spain. KAGRA is supported by Ministry of Education, Culture,
Sports, Science and Technology (MEXT), Japan Society for the Promotion of
Science (JSPS) in Japan; National Research Foundation (NRF) and Ministry of
Science and ICT (MSIT) in Korea; Academia Sinica (AS) and National Science and
Technology Council (NSTC) in Taiwan.
This research was supported in part by NSF grant 2012857 to the University of
California at San Diego, NSF grants 1912625 and 2309305 to Texas Tech
University, and NSF grant 2110460 to the Rochester Institute of Technology.
The authors acknowledge computational resources provided by the High Performance
Computing Center (HPCC) of Texas Tech University at Lubbock
(http://www.depts.ttu.edu/hpcc/).
This paper has LIGO Document number P2300319.

\end{acknowledgments}

\bibliography{paper}

\end{document}